\documentstyle[preprint,tighten,aps]{revtex} 
\begin{document}           %
\draft
\preprint{\vbox{\noindent
\null \hfill hep-ph/9811352\\
          \null\hfill  INFNCA-TH9815}}
\title{Vacuum oscillations and the distorted solar neutrino spectrum 
observed by Superkamiokande
      }
\author{V. Berezinsky$^{1,}$\cite{email1},
         G.~Fiorentini$^{2,}$\cite{email2},
         and M.~Lissia$^{3,}$\cite{email3}}
\address{
$^{1}$ Istituto Nazionale di Fisica Nucleare, Laboratori Nazionali del
       Gran Sasso, SS. 16 bis, I-67010 Assergi (AQ), Italy\\
       and Institute for Nuclear Research, Moscow, Russia \\
$^{2}$ Dipartimento di Fisica dell'Universit\`a di Ferrara and Istituto
      Nazionale di Fisica Nucleare, Sezione di Ferrara, 
      via Paradiso 12, I-44100 Ferrara, Italy\\
$^{3}$ Istituto Nazionale di Fisica Nucleare, Sezione di Cagliari and
      Dipartimento di Fisica dell'Universit\`a di Cagliari,
      I-09042 Monserrato (CA), Italy}
%
\date{November 24, 1998}
\maketitle                 
\begin{abstract}
The excess of solar-neutrino events above 13~MeV that has been recently
observed by Superkamiokande can be explained by vacuum oscillations (VO).
If the boron neutrino flux is 20\% smaller than the standard solar
model (SSM) prediction and the chlorine signal is assumed 30\% 
(or $3.5 \sigma$) higher
than the measured one, there exists a VO solution that reproduces both
the observed boron neutrino spectrum, including the high energy distortion,
and the other measured neutrino rates.
This solution is already testable by the predicted
anomalous seasonal variation of the gallium signal. Its most distinct
signature, a large anomalous seasonal variation of $^7$Be neutrino flux,
can be easily observed by the future detectors, BOREXINO and LENS.
\end{abstract}
%
%
\narrowtext
\section*{}
Superkamiokande~\cite{REF_SKAM} has recently observed
an excess of solar-neutrino events at energy higher than $13~$MeV.
This excess cannot be interpreted as a distortion of the boron
neutrino spectrum due to neutrino oscillations~\cite{REF_SKAM,REF_BKS},
if one restricts oneself to the standard oscillation solutions that
explain the observed gallium, chlorine and water-cerenkov neutrino rates.

It is tempting to think that this excess is a result of low statistics 
at the end of the boron neutrino spectrum. 
On the other hand, all systematic errors are reduced at higher energies:
the background decreases, the detection efficiency increases, and
the recoil electron direction is better determined. The data from 
the SNO detector, which is in the operation now ({\em e.g.},
see~\cite{REF_SNO}), can shed light on this excess.

Another possible explanation~\cite{escri,REF_BaKr} is that the 
hep neutrino flux might be significantly larger (about a factor 20--30) 
than the
SSM prediction. The hep flux depends on solar properties, such as the
$^3$He abundance and  the temperature, and on $S_{13}$, the zero-energy 
astrophysical $S$-factor of the 
$p+{}^3\text{He} \rightarrow {}^4\text{He} + e^+ + \nu$ reaction.
Both SSM based~\cite{REF_BaKr} and model-independent~\cite{REF_BDFR}
approaches give a robust prediction for ratio
$\Phi_{\nu}(\text{hep})/S_{13}$.
Therefore, this scenario implies a cross-section larger by
a factor 20--30  than the present calculations (for
reviews see~\cite{REF_BaKr,INT}). Such a huge mistake in the calculation
does not seem likely, though the required large cross-section does not 
contradict to the ``first principle physics''~\cite{REF_BaKr}.

In this letter we propose another explanation of the observed 
excess based on the distortion of spectrum by vacuum oscillations. 

Combined with the SSM, VO can explain the observed rates of all 
three solar neutrino experiments (for reviews
see~\cite{REF_Bahc,REF_BiPe,REF_Turck,REF_Hax}).
We shall refer to these models
as Standard Vacuum Oscillation (SVO) solutions. A recent detailed
study~\cite{REF_HaLa95,REF_BK96,REF_KP96,REF_HaLa97,REF_Lisi} of SVO solutions 
shows that global fits to the data
result in oscillation parameters 
within the ranges 
$5\cdot 10^{-11}~\text{eV}^2\leq \Delta m^2 
\leq 1\cdot 10^{-10}~\text{eV}^2$ 
and $0.7\leq\sin^2 2\theta\leq 1$ for oscillations between the active 
neutrino components.
A large range for $\sin^2 2\theta$ is caused by uncertainties in the 
B-neutrino flux. 
This effect has been explicitly investigated in Refs.~\cite{REF_KrSm,REF_KP96}.
In the SSM the B-neutrino flux uncertainties
($+19\%,-14\%, $~\cite{REF_BP98}) are caused by the uncertainties
in  $S_{17}$ (the $p$-Be cross section is poorly known)
and by the strong temperature dependence of this flux.
The above uncertainties are given for $1\sigma$ errors and they could be 
larger, especially due to
the $S_{17}$ factor. Motivated by it, several authors
considered the boron flux as $\Phi_B=f_B\Phi_B^{SSM}$ with
$f_B$ as a free parameter~\cite{REF_BKS,REF_KrSm,REF_HaLa95,REF_KP96}. 

A signature of VO is the anomalous seasonal variation 
of the neutrino flux~\cite{REF_Pom,REF_BiPe}. The variation of the distance 
between the Sun
and the Earth affects the detected flux, apart from a trivial geometrical
factor, because of the dependence of the survival probability
$P(\nu_e \to \nu_e)$ on the distance. This effect is absent for
the MSW solutions.
The anomalous seasonal variation is the strongest for the Be-neutrino
flux~\cite{REF_BiPo,REF_GK,REF_KrPe95,REF_Lisi}.  

The seasonal variations 
for $\Delta m^2$ larger than the values allowed by the SVO solutions were 
analyzed recently in Ref.~\cite{REF_GKK}. The authors found some significant 
consequences such as energy dependence and correlation with distortion of 
the spectrum. The latter effect was also discussed earlier in
Ref.~\cite{REF_MS}. A clear discussion of the 
seasonal variation effect has been presented in Ref.~\cite{REF_Rosen}

To explain the distortion of spectrum observed in Superkamiokande
we allow a boron neutrino flux 15--20\% smaller than the SSM 
prediction, and we allow that the chlorine signal be about 30\%
larger than the Homestake observation.
This assumed $3.5\sigma$ increase could have a statistical origin or 
might imply a small systematic error in the Homestake 
experiment, though we do not have any concrete argument in favor of such 
``theoretical assumption''. 

In our calculations, we shall use neutrino fluxes from the BP98
model~\cite{REF_BP98} with the B-neutrino flux rescaled as 
$\Phi_B= f_B \Phi_B^{SSM}$. 
For the chlorine rate we adopt the recent Homestake data~\cite{REF_Hom}
multiplied by a factor $f_{Cl}$: $2.56f_{Cl}\pm 0.16 \pm 0.15$~SNU. For the 
gallium
rate we use the average of the GALLEX~\cite{REF_Ki} and SAGE~\cite{REF_Ga} 
results:
$72.5 \pm 5.7$~SNU. Finally, we take the Superkamiokande
result~\cite{REF_SKAM}:
$(2.46 \pm 0.09)\cdot 10^6$~cm$^{-2}$s$^{-1}$.
For each pair $f_B$ and $f_{Cl}$
we find the VO solution, {\em i.e.}, the parameters
($\Delta m^2, \sin^22\theta$), that explains
the observed rates, and then we calculate the corresponding 
boron neutrino spectrum.

For example, for $f_B=0.8$ and $f_{Cl}=1.3$ the oscillation 
parameters ($\Delta m^2=4.2 \cdot 10^{-10}$~eV$^2, \sin^22\theta=0.93$)
give a good fit to all rates ($\chi^2$/d.o.f. = 3.0/3). On the other hand,
the same oscillation parameters give a good fit~\cite{REF_SKAM} 
to the distorted Superkamiokande spectrum.
More generally,
this choice of oscillation parameters gives rates in agreement with the
experiments for $0.77 \leq f_B \leq 0.83$ and $1.3 \leq f_{Cl} \leq 1.55$.

In Fig.~1 we present the spectra of the VO solutions as the ratio to the
SSM unmodified spectrum~\cite{REF_BP98}. The dotted and dashed curves show
two spectra corresponding to the SVO solutions of Ref.~\cite{REF_BKS} and
Ref.~\cite{REF_HaLa97}, respectively. The solid line shows the
spectrum-distorted vacuum oscillation (DVO) solution that corresponds to
$\Delta m^2=4.2\cdot 10^{-10}$~eV$^2$ and $\sin^2 2\theta=0.93$  
($f_B=0.8$ and $f_{Cl}=1.3$). The DVO spectrum differs from SVO 
at low ($E \sim 5 - 6~MeV$) and high ($E>13~MeV$) energies. Both 
deviations can be tested by future Superkamiokande and SNO data.

 The role of the two parameters, $f_B$ and $f_{Cl}$, for the best fit 
of the spectrum is
different: while $f_B$ mostly changes $\sin^2 2\theta$, $f_{Cl}$
affects $\Delta m^2$ and, therefore, the spectrum. Values of $f_{Cl}$
as low as 1.2 already give bad fit to the observed spectrum.   

The anomalous seasonal variations of Be-neutrino flux and of the gallium
signal are shown in the Fig.~2. Anomalous seasonal 
variation is described by the survival probability of the electron neutrino 
$P(\nu_e \to \nu_e)$. For Be-neutrinos with energy $E=0.862$~MeV 
the survival probability (the suppression factor) is given by
\begin{equation}
\label{EQ1}
P(\nu_e \to \nu_e)= 1-\sin^2 2\theta
\sin^2\left( \frac{\Delta m^2 a}{4E}\, (1+e\cos \frac{2\pi t}{T} ) \, \right)
\, ,
\end{equation}
where $a=1.496\cdot 10^{13}$~cm is the semimajor axis, 
$e=0.01675$ is the eccentricity of the Earth's orbit, and $T=1$~yr is the 
orbital period. The phase in 
Eq.~(\ref{EQ1}) is such that $t=0$ corresponds to the aphelion.
In Fig.~2 the solid and dashed curves show the variation of the Be-neutrino
flux for the DVO and SVO~\cite{REF_BKS} cases, respectively. The case of
the DVO (solid curve) is dramatically different from the SVO case: there are 
two maxima and minima during one year and the survival probability oscillates 
between 
$1-\sin^2 2\theta \approx 0.14$ and 1. The explanation is obvious: the DVO 
solution has a large $\Delta m^2$, which results in a phase
$\Delta m^2 a/(4E) \approx 93$, large enough to produce two full harmonics 
during one year, when the phase changes by about 3\% due to the factor
$(1+e\cos 2\pi t/T)$. 
The flat central maximum with a shallow local minimum has
a trivial trigonometric explanation (for some parameters this accidental
shallow minimum turns into maximum). The phases of maxima and minima 
are not fixed in the DVO solution, because tiny changes of
$\Delta m^2$ shift their positions.
 
Therefore, the DVO solution predicts that the beryllium electron neutrinos
should arrive almost unsuppressed during about four months in a year!

According to the SSM, beryllium neutrinos contribute 34.4~SNU out of
the total gallium signal of 129~SNU.
Therefore, the strong $^7$Be neutrino oscillation
predicted by the DVO solution also implies an appreciable variation of 
total gallium signal. In Fig.~2 the dotted curve shows this variation
corresponding to the DVO solution, which can be compared with the
weaker variation corresponding to the SVO solution (dashed-dotted curve).
It is possible that the DVO variation could already be tested
by the existing gallium data, and this possibility will significantly
increase when the new results from GNO with enlarged mass are available.

In Fig.~3 the predicted time variation of the gallium signal is compared 
with GALLEX data (M~.Cribier cited in \cite{REF_Ki}). The data give the 
rates averaged for the same two months every year of observations. 
The theoretical prediction (solid curve) is plotted with the same averaging. 
The $7\%$ geometrical variation is included with the proper phase. The 
phase of the oscillation and the averaged flux have been chosen to fit 
the data. The fit by the theoretical curve has $\chi^2$/d.o.f. = 0.85/4; the 
fit by a time-independent signal is also good: $\chi^2$/d.o.f. = 1.36/5.
Because  of limited statistics, we 
do not interpret the good agreement seen in Fig.~3 as a proof of DVO 
solution, though one might consider it as some indication. 

The anomalous seasonal variation of Be-neutrino flux predicted by the DVO 
solution can be reliably observed by the future
BOREXINO~\cite{REF_BOREXINO} and LENS~\cite{REF_LENS} detectors. 
Additionally, LENS, which should measure the flux and spectrum of $pp$ 
neutrinos, will be able to observe the suppression of $pp$
neutrino flux, $P(\nu_e \to \nu_e)= 1- (1/2)\sin^2 2\theta=0.53$,
which is another signature of VO solutions.

In conclusion, a B-neutrino flux 20\% lower than in the SSM
(easily allowed by the present uncertainties) and the assumption of
chlorine signal $30 \%$ ($3.5\sigma$) higher than the Homestake data 
result in a VO
solution with a distorted neutrino spectrum that fits the one
recently observed by Superkamiokande. This solution predicts strong
seasonal variation of $^7$Be-neutrino flux, which would be seen by future 
experiments, and appreciable gallium-signal variation, which is compared 
in Fig.~3 with existing data. 

\acknowledgments
We are grateful to A.~Bettini and A.~Yu.~Smirnov for useful discussions.

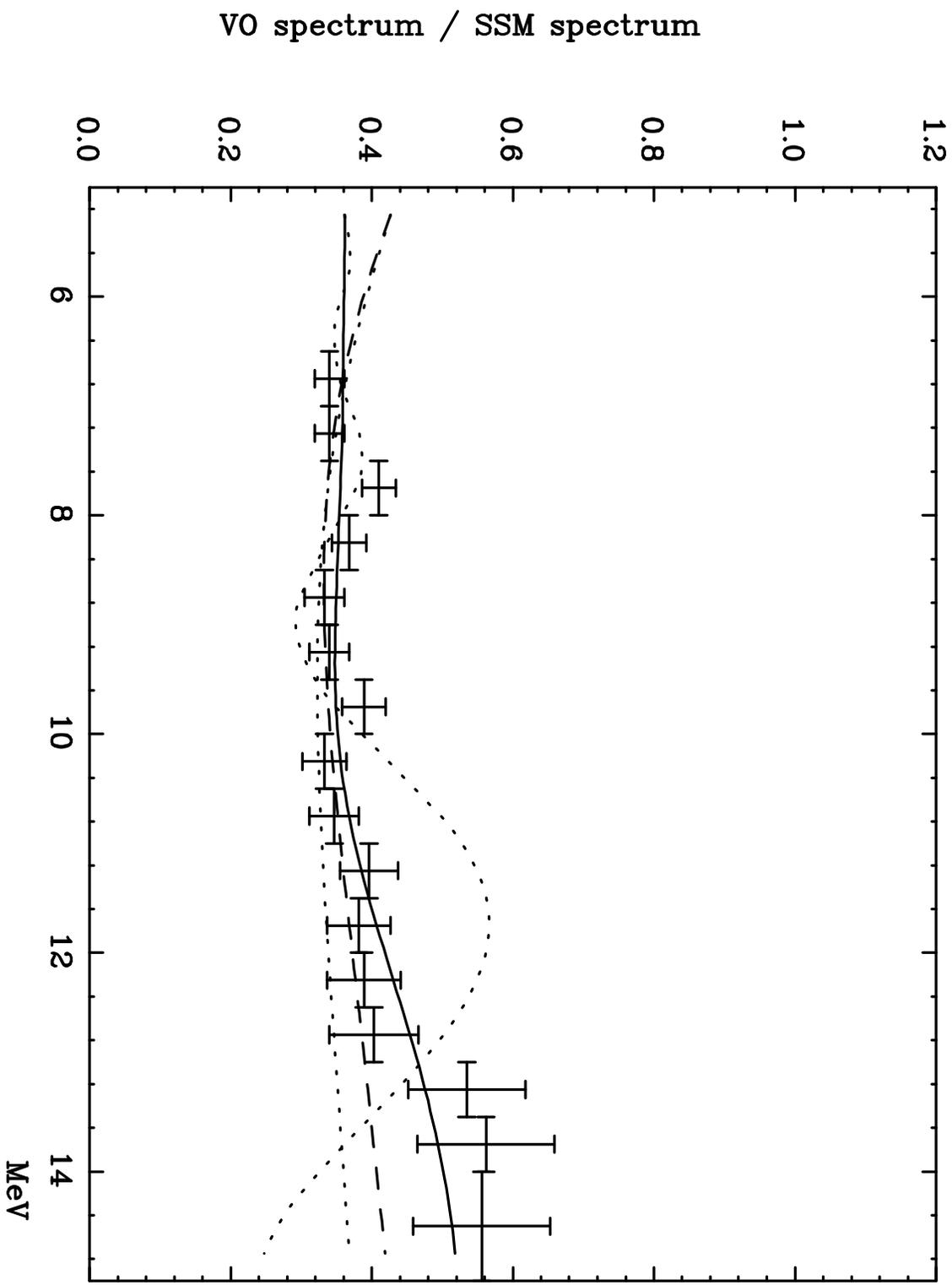
\begin{figure}
\caption[a]{
Ratio of the VO spectrum to the SSM spectrum.
The solid curve corresponds to the DVO solution with
$\Delta m^2=4.2\cdot 10^{-10}$~eV$^2$ and $\sin^2 2\theta=0.93$.
The dashed and dotted curves correspond to the SVO solutions
of Refs.~\cite{REF_HaLa97} and \cite{REF_BKS}, respectively.
Energy resolution is taken into account everywhere.
The data points show the Superkamiokande results~\cite{REF_SKAM}.
}
\end{figure}

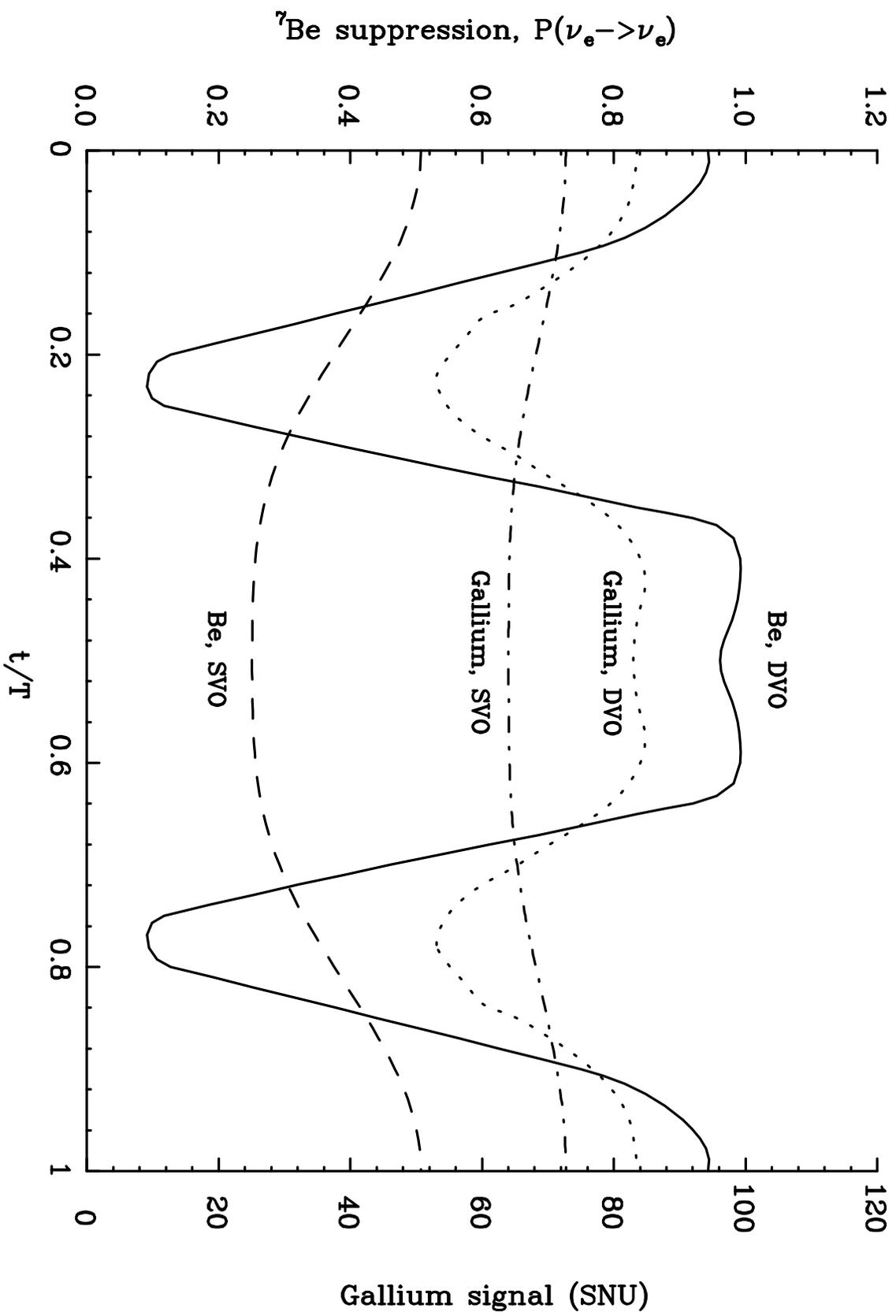
\begin{figure}
\caption[b]{
Anomalous seasonal variations of the beryllium neutrino flux and gallium
signal for the SVO and DVO solutions.
The survival probability $P(\nu_e\to \nu_e)$ for Be neutrinos (suppression 
factor) is given for the DVO (solid curve) and the SVO (dashed curve) 
solutions as function of time ($T$ is an orbital period). The dotted
(dash-dotted) curve shows the time variation of gallium signal in SNU for
the DVO (SVO~\cite{REF_BKS}) solution.
           }
\end{figure}
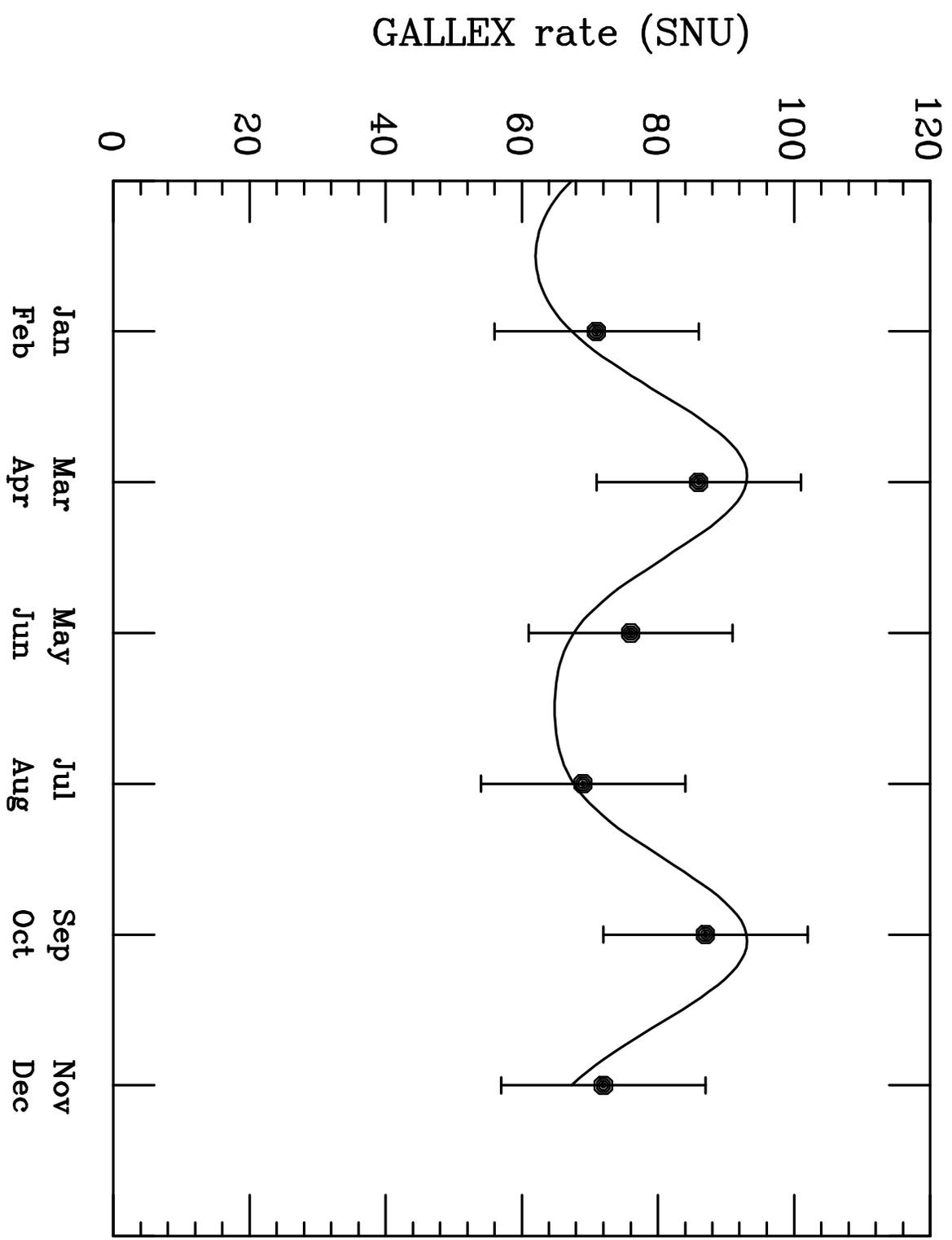
\begin{figure}
\caption[c]{
Seasonal time variation predicted by the DVO solution in comparison with 
the GALLEX data. The theoretical dependence includes oscillations and $7\%$ 
geometrical       
variations. The phase of the oscillation (undefined in DVO) and 
the mean rate (chosen in DVO as averaged rate of GALLEX and SAGE) has been
chosen here to fit the data. The fit by the DVO solution gives
$\chi^2$/d.o.f. = 0.85/4 and the fit by a time-independent signal gives
$\chi^2$/d.o.f. = 1.36/5.
           }
\end{figure}

\end{document}